\documentclass{article}
\usepackage{graphicx}
\usepackage{amsmath}
\usepackage{amsfonts}
\usepackage{amssymb}
\newtheorem{theorem}{Theorem}
\newtheorem{acknowledgement}[theorem]{Acknowledgement}

\begin{document}

\title{Magnetoroton scattering by phonons in the fractional quantum Hall r\'{e}gime}
\author{KA Benedict and RK HIlls\\School of Physics and Astronomy, \\The University of Nottingham\\University Park\\NOTTINGHAM\\NG7 2RD\\U.K.}
\date{}
\maketitle
\begin{abstract}
Motivated by recent phonon spectroscopy experiments in the fractional quantum
Hall regime we consider processes in which thermally excited magnetoroton
excitations are scattered by low energy phonons. We show that such scattering
processes can never give rise to dissociation of magnetorotons into unbound
charged quasiparticles as had been proposed previously. In addition we show
that scattering of magnetorotons to longer wavelengths by phonon absorption is
possible because of the shape of the magnetoroton dispersion curve and it is
shown that there is a characteristic cross-over temperature above which the
rate of energy transfer to the electron gas changes from an exponential
(activated) to a power law dependence on the effective phonon temperature.
\end{abstract}

\section{Introduction}

Phonon spectroscopy is a technique that has been used with much success in the
study of semiconductor systems\cite{gen-phon} and in particular the properties
of two-dimensional electron systems (2des) formed at heterojunctions and
quantum wells. One of the most important recent applications of the technique
has been to the study of the incompressible quantum liquid of electrons formed
at a semiconductor heterojunction which gives rise to the fractional quantum
Hall effect\cite{PRL1,PRL2}. The essential points of the experiments are that
a pulse of phonons with an approximately black body spectrum of energies is
injected into the substrate of a semiconductor device. The phonons propagate
ballistically across the substrate until they encounter the electrons at the
heterojunction where some of them may be absorbed. In the fractional quantum
Hall states the electrons have a true energy gap as a consequence of
incompressibility, so that if the electrons are initially at zero temperature,
only the phonons in the high energy part of the black body distribution can be
absorbed. It is a great virtue of the technique that these phonons may also
have wave vectors comparable with the interesting structure in the dispersion
relation predicted for the low lying collective modes or \emph{magnetorotons
}\cite{GMP} of the electron liquid. The basic theoretical picture of what
happens in these experiments has been outlined recently \cite{Theory}. This
work concentrated on processes in which the absorption of a high energy phonon
leads to the creation of a magnetoroton excitation close to the minimum in the
dispersion curve. In the experiments, the effect of the phonon absorption is
detected by measuring changes in the two-terminal resistance of the device. It
was speculated \cite{PRL1} that in some way the elementary excitations created
by the absorption of phonons contributed more or less directly to the
increased dissipation observed in the two-terminal measurements. Dissipation
due to mutual friction between the magnetoroton gas and the Laughlin liquid
has been discussed by Platzman \cite{Platzman}. A more concrete mechanism for
dissipation due to magnetoroton creation was proposed by one of us
\cite{Wurtzburg}. The essential idea of this was that, once created by
absorption of a high-energy phonon, a magnetoroton excitation would rapidly
absorb many low-energy phonons from the bulk of the Planck distribution in the
pulse. For wave vectors larger than $q^{\ast}$ the magnetorotons can be viewed
as excitons formed from Laughlin quasiparticles and quasiholes \cite{Laughlin}%
, the separation of the two quasiparticles in an exciton with wave vector $q$
being $l_{c}^{2}q$. Once an exciton has absorbed sufficient low energy phonons
that the separation of its constituents exceeds some effective screening
length, the fractionally charged quasiparticles are effectively unbound and
contribute directly to the two-terminal resistance. A crude analysis presented
in \cite{Wurtzburg} indicated that this process might not be particularly
rapid. One purpose of this paper is to analyse this idea more carefully and in
fact to show that the process described will not actually happen at all. The
shape of the magnetoroton dispersion curve is such that the processes in which
magnetorotons absorb energy from the low-energy rump of the Planck
distribution will always reduce the wave vector of the magnetoroton. Hence we
conclude, as discussed in \cite{Theory}, that the energy absorbed from the
phonon pulse by the electrons is shared among all the electrons leading to an
overall increase in the electron temperature, which then leads to an increased
two-terminal resistance for the device. The processes described above in which
magnetorotons are heated with a corresponding decrease in wave vector will, of
course, occur in real experiments and will provide the leading correction to
the expressions described in \cite{Theory} at finite electron temperature.
These corrections are of particular importance for recent experiments
\cite{PRL2} in which the time dependence of the two-terminal resistance, and
therefore the electron temperature, over the duration of the phonon pulse is
well resolved. The layout of this paper is as follows. In section two a brief
review of the properties of electrons in the fractional quantum Hall state and
the phonons in the substrate is given, more details can be found in \cite{GMP}
and \cite{Theory}. In section 3 expressions for the rate of energy transfer
between a phonon pulse and the electron system will be given and the forms of
the relevant matrix elements discussed. In sections four a simple analytical
treatment of these rate expressions will be given which shows that
magnetoroton dissociation due to heating by low energy phonons cannot occur.
In section 5 the contribution of the magnetoroton scattering processes to the
rate of energy transfer from the pulse to the 2des will be considered and it
will be shown that there is a (low) cross-over temperature at which the
dependence of the transfer rate on heater temperature changes from exponential
to a power law. Section 6 will contain a brief summary and discussion.

\section{Electrons and Phonons in Fractional Quantum Hall Systems}

Our basic picture of the physics of electrons in the fractional quantum Hall
states emerges from the ideas of Laughlin \cite{Laughlin} and Girvin,
MacDonald and Platzman \cite{GMP}. Laughlin constructed the wave function for
a state with exceptionally low energy at the primary quantum Hall filling
factor $\nu=1/\left(  2m+1\right)  $. He also constructed low energy
quasiparticle excitations which carry fractional charge and have an energy gap
even in the thermodynamic limit. Hence the description of the quantum Hall
state as an incompressible quantum liquid. Girvin, MacDonald and Platzman
\cite{GMP} proposed wave functions for a branch of neutral collective
excitations whose energy lies below that of the quasiparticle continuum. This
situation is analogous to the phonon-roton branch in superfluid helium. GMP
made this analogy more explicit by adapting the theoretical approach of
Feynman \cite{Feynman} to the case of magnetically quantized 2d electrons.
They found a branch of low energy excitations whose dispersion relation,
$\Delta\left(  q\right)  $, has a deep minimum at a finite wave vector,
$q^{\ast}$. By analogy with liquid helium they dubbed these magnetorotons.
Unlike helium, the fractional quantum Hall liquid does not have a gapless
phonon mode at long wavelengths, the incompressibility of the underlying state
causes the energy of the collective mode to increase for wavelengths longer
than the magnetoroton minimum and to always be gapped. One expects the single
mode approximation of GMP to give a good quantitative account of the physics
of this collective mode for wavelengths close to and longer than the
magnetoroton minimum. However, for shorter wavelengths one expects the mode to
be better described as an exciton composed of a pair of oppositely charged
Laughlin quasiparticles. The separation of the pair in a state of wave vector
$q$ being $l_{c}^{2}q/\nu$. The form of the dispersion relation which we
assume here is
\begin{equation}
\Delta\left(  q\right)  =\left\{
\begin{array}
[c]{cc}%
\Delta_{\text{SMA}}\left(  q\right)  & q<q_{m}\\
\Delta_{\infty}-E_{c}\frac{\nu^{3}}{q} & q>q_{m}%
\end{array}
\right. \label{Dispersion}%
\end{equation}
where $E_{c}=e^{2}/4\pi\varepsilon_{0}\kappa l_{c}$ is the Coulomb energy
scale, $\Delta_{\text{SMA}}$ is the dispersion relation obtained from the
single mode approximation of GMP \cite{GMP} and $q_{m}$ and $\Delta_{\infty}$
are chosen so that $\Delta\left(  q\right)  $ and its first derivative are
continuous for all $q$. This is shown in figure 1 below. As can be seen the
``exciton'' part of this dispersion relation is very flat: a fact that will
have considerable significance for the proposed dissociation mechanism.

Phonons in a GaAs substrate travel ballistically for frequencies up to about a
terahertz. Hence a pulse of phonons injected into a GaAs substrate will move
across the substrate of a heterostructure until it encounters the
heterojunction where it will interact with the vertically confined electrons.
The Hamiltonian for the electron phonon interaction can be written in the
form
\begin{equation}
H_{e\phi}=\sum_{s,\underline{Q}}M_{s}\left(  \underline{Q}\right)  Z\left(
q_{z}\right)  \widehat{\rho}_{\mathbf{q}}\left(  \widehat{a}_{s}\left(
\underline{Q}\right)  +\widehat{a}_{s}^{\dagger}\left(  -\underline{Q}\right)
\right)
\end{equation}
where $\widehat{\rho}_{\mathbf{q}}$ is the Fourier transform of the local (2d)
electron density, $\widehat{a}_{s}\left(  \underline{Q}\right)  $ is the
second quantized operator for phonons of polarization $s$ and (3d) wave vector
$\underline{Q}$, $Z\left(  q_{z}\right)  $ is a form factor to account for the
finite thickness of the electron layer and $M_{s}\left(  \underline{Q}\right)
$ is the matrix element for the coupling of phonons in the given mode to
electrons close to the $\Gamma$ point of the GaAs conduction band. This
hamiltonian incorporates both piezoelectric and deformation potential coupling
which differ in their anisotropy and in the index of the power law dependence
on phonon energy.

The trial wave function used by GMP for a magnetoroton of wave vector
$\mathbf{q}$ is of the form
\begin{equation}
\left|  \mathbf{q}\right\rangle =\frac{\overline{\rho}_{\mathbf{q}}\left|
\Psi_{L}\right\rangle }{\sqrt{\left\langle \Psi_{L}\right|  \overline{\rho
}_{-\mathbf{q}}\overline{\rho}_{\mathbf{q}}\left|  \Psi_{L}\right\rangle }}%
\end{equation}
where $\left|  \Psi_{L}\right\rangle $ is the Laughlin ground state and
$\overline{\rho}_{\mathbf{q}}$ is the projection of the electron density
operator onto the lowest Landau level: $\overline{\rho}_{\mathbf{q}%
}=\mathcal{P}_{0}\rho_{\mathbf{q}}\mathcal{P}_{0}$ (see \cite{Jachs} for
details of the projection). If we consider a pulse of phonons characterized by
a distribution function $n_{s}\left(  \underline{Q}\right)  $ in contact with
a 2des in the fractional quantum Hall state at zero temperature, then the rate
of energy transfer to the electrons should be, using Fermi's golden rule
\begin{align}
P  & =2\pi\sum_{s,Q,\mathbf{k}}\left|  M_{s}\left(  \underline{Q}\right)
Z\left(  q_{z}\right)  \left\langle \mathbf{q}\right|  \widehat{\rho
}_{\mathbf{q}}\left|  \Psi_{L}\right\rangle \right|  ^{2}\delta\left(
\omega_{s}\left(  \underline{Q}\right)  -\Delta\left(  \mathbf{k}\right)
\right)  n_{s}\left(  \underline{Q}\right) \\
& =2\pi N\sum_{s,\underline{Q},\mathbf{k}}\left|  M_{s}\left(  \underline
{Q}\right)  Z\left(  q_{z}\right)  {}\right|  ^{2}\overline{s}\left(
q\right)  \delta\left(  \omega_{s}\left(  \underline{Q}\right)  -\Delta\left(
\mathbf{k}\right)  \right)  n_{s}\left(  \underline{Q}\right) \label{creation}%
\end{align}
\qquad where
\begin{equation}
\overline{s}\left(  q\right)  =\frac{1}{N}\left\langle \Psi_{L}\right|
\overline{\rho}_{-\mathbf{q}}\overline{\rho}_{\mathbf{q}}\left|  \Psi
_{L}\right\rangle
\end{equation}
is the projected static structure factor of the electron liquid and we set
$\hbar=l_{c}=1$ from now on. The projected structure factor is related, as
shown by GMP, to the full structure factor, $s\left(  q\right)  $ by
\begin{equation}
\overline{s}\left(  q\right)  -e^{-q^{2}/2}=s\left(  q\right)  -1\qquad.
\end{equation}
GMP extracted $s\left(  q\right)  $ from Laughlin's wave function and used it
in the calculation of the magnetoroton dispersion $\Delta\left(  q\right)  $.
\ The expression in equation \ref{creation} forms the basis for the work in
\cite{Theory}.

\section{Magnetoroton-Phonon Scattering}

Let us now consider the probability per unit time that a magnetoroton of wave
vector $\mathbf{k}$ will absorb a phonon with wave vector $\underline
{Q}=\left(  \mathbf{q},q_{z}\right)  $ and be scattered into the state with
wave vector $\mathbf{k+q}$. Fermi's golden rule tells us that this will be
\begin{equation}
\tau_{\mathbf{k,}\underline{Q}}^{-1}=2\pi\left|  M_{s}\left(  \underline
{Q}\right)  Z\left(  q_{z}\right)  \right|  ^{2}\left|  \left\langle
\mathbf{k+q}\right|  \rho_{\mathbf{q}}\left|  \mathbf{k}\right\rangle \right|
^{2}\delta\left(  \Delta\left(  k\right)  +\omega_{s}\left(  \underline
{Q}\right)  -\Delta\left(  \left|  \mathbf{k+q}\right|  \right)  \right)
\end{equation}
which we can express, using the ansatz of GMP, as
\begin{equation}
2\pi\frac{\left|  M_{s}\left(  \underline{Q}\right)  Z\left(  q_{z}\right)
\right|  ^{2}}{\overline{s}\left(  k\right)  \overline{s}\left(  \left|
\mathbf{k+q}\right|  \right)  }\left|  P^{\left(  3\right)  }\left(
\mathbf{k,q}\right)  \right|  ^{2}\delta\left(  \Delta\left(  k\right)
+\omega_{s}\left(  \underline{Q}\right)  -\Delta\left(  \left|  \mathbf{k+q}%
\right|  \right)  \right)  \qquad,
\end{equation}
where we define the projected three-point correlation function
\begin{equation}
P^{\left(  3\right)  }\left(  \mathbf{k,q}\right)  =\frac{1}{N}\left\langle
\Psi_{L}\right|  \overline{\rho}_{\mathbf{-k-q}}\overline{\rho}_{\mathbf{q}%
}\overline{\rho}_{\mathbf{k}}\left|  \Psi_{L}\right\rangle \qquad.
\end{equation}
The projected three-point correlation function has been discussed by MacDonald
et al. \cite{MacDonald} in the context of magnetoroton scattering by quenched
impurities. As outlined there, this quantity can be related to its unprojected
counterpart via
\begin{align}
P^{\left(  3\right)  }\left(  \mathbf{k,q}\right)   & =e^{-\frac{1}{2}%
q^{2}-\frac{1}{2}\left(  \mathbf{k,q}\right)  }\left\{  e^{-\frac{1}{2}k^{2}%
}+s\left(  k\right)  -1\right\}  +e^{-\frac{1}{2}\left(  \left(
\mathbf{k+q}\right)  ,\mathbf{k}\right)  }\left(  s\left(  q\right)  -1\right)
\label{P3}\\
& +e^{\frac{1}{2}\left(  \mathbf{q},\mathbf{k}\right)  }\left(  s\left(
\left|  \mathbf{k+q}\right|  \right)  -1\right)  +n^{(3)}\left(
\mathbf{k,q}\right) \nonumber
\end{align}
where
\begin{align}
\left(  \mathbf{k,q}\right)   & =\mathbf{k\cdot q+}i\mathbf{k\wedge q}\\
& =\left(  k_{x}-ik_{y}\right)  \left(  q_{x}+iq_{y}\right) \nonumber
\end{align}
and
\begin{equation}
n^{(3)}\left(  \mathbf{k,q}\right)  =\frac{1}{N}\left\langle \Psi_{L}\right|
\rho_{\mathbf{-k-q}}\rho_{\mathbf{q}}\rho_{\mathbf{k}}\left|  \Psi
_{L}\right\rangle \qquad.
\end{equation}
MacDonald et al. used the convolution approximation \cite{Jackson} developed
for the study of corresponding processes in superfluid helium to estimate this
latter quantity in terms of the static structure factor and obtained an
expression for $n^{\left(  3\right)  }\left(  \mathbf{k,q}\right)  $ of the
form
\begin{align}
n^{\left(  3\right)  }\left(  \mathbf{k,q}\right)   &  =N^{2}\delta
_{\mathbf{k,0}}\delta_{\mathbf{q,0}}+N\delta_{\mathbf{k,0}}h\left(  q\right)
+N\delta_{\mathbf{q,0}}h\left(  k\right)  +N\delta_{\mathbf{k+q,0}}h\left(
k\right) \label{Convolution}\\
&  +h\left(  k\right)  h\left(  q\right)  +\left[  h\left(  k\right)
+h\left(  q\right)  +h\left(  k\right)  h\left(  q\right)  \right]  h\left(
\left|  \mathbf{k+q}\right|  \right)  \qquad\nonumber
\end{align}
where
\begin{equation}
h\left(  k\right)  =s\left(  k\right)  -1-N\delta_{\mathbf{k,0}}\qquad.
\end{equation}
Beyond this, very little is known about this function, although a useful
sum-rule for testing approximation schemes for it was recently derived in
\cite{Brownlie}. Although more detailed knowledge of this quantity would be
desirable for truly quantitative work on the phonon experiments, here we will
restrict ourselves to the more modest objective of qualitative understanding.

\section{Analytic Estimates}

We will begin by deriving a simple criterion for small momentum transfer
magnetoroton scattering processes. Energy and momentum conservation require
that a magnetoroton with initial wave vector $\mathbf{k}$ can absorb a phonon
with in-plane momentum $\mathbf{q}$ incident at angle $\theta$ to the normal
to the 2des only when
\begin{equation}
\Delta\left(  \left|  \mathbf{k+q}\right|  \right)  =\Delta\left(  k\right)
+\frac{c_{s}q}{\sin\theta}\qquad.
\end{equation}
For small $q$ we set
\begin{equation}
\Delta\left(  \left|  \mathbf{k+q}\right|  \right)  \approx\Delta\left(
k\right)  +\frac{\mathbf{q\cdot k}}{k}\Delta^{\prime}\left(  k\right)
\end{equation}
so that
\begin{equation}
\frac{\mathbf{q\cdot k}}{kq}=\cos\phi=\frac{c_{s}}{\Delta^{\prime}\left(
k\right)  \sin\theta}%
\end{equation}
where $\cos\phi=\mathbf{k\cdot q}/kq$. This can only be satisfied when
\begin{equation}
\left|  \Delta^{\prime}\left(  k\right)  \right|  >\frac{c_{s}}{\sin\theta}%
\end{equation}
This result simply means that phonon absorption can only occur when the slope
of the magnetoroton dispersion is greater than the slope of the dispersion
curve for phonons with an angle of incidence $\theta$. By happy accident in
our system of units ($\hbar=l_{c}=\mathcal{E}_{c}=1$) \ the unit of velocity,
$l_{c}\mathcal{E}_{0}/\hbar$, is independent of magnetic field and, for GaAs,
has the value $1.68\times10^{5}%
\operatorname{m}%
\operatorname{s}%
^{-1}$. In these units the speeds of sound are $c_{1}\approx0.0305$ for
LA\ phonons and $c_{2}=c_{3}\approx0.0196$ for TA\ phonons. Hence we expect
that these scattering processes will be suppressed at large $k$ and
sufficiently close to the magnetoroton minimum where the magnetoroton
dispersion is very flat and $\Delta^{\prime}\left(  k\right)  $ small.

Let us estimate where this suppression occurs for large $k$ to assess the
viability of the magnetoroton dissociation idea. At large $k$ we assume
\begin{equation}
\Delta\left(  k\right)  =\Delta_{\infty}-\frac{\nu^{3}}{k}%
\end{equation}
so that
\begin{equation}
\Delta^{\prime}\left(  k\right)  =\frac{\nu^{3}}{k^{2}}%
\end{equation}
and scattering will be cut-off beyond
\begin{equation}
k_{s}=\sqrt{\frac{\nu^{3}\sin\theta}{c_{s}}}\qquad.
\end{equation}
Hence we have, for $\nu=1/3$, $l_{c}k_{1}=1.102$, $l_{c}k_{2}=1.375$, assuming
that phonons with $\theta=\pi/2$ (i.e. in plane phonons) exist in the pulse.
The latter cut-off is only just beyond the magnetoroton minimum while the
former is so small that the whole analysis is entirely meaningless. In fact,
in the experiments, phonons with $\theta=\pi/2$ are strongly focussed away
from the device: only phonons with $\theta<\pi/4$ are likely to couple with
the electrons ensuring that there will be no coupling to very low energy
phonons at all in the ``exciton'' regime. The corrections to this result
arising from finite values of the phonon wave vector only strengthen it: the
curvature of the magneto-exciton dispersion curve makes it even harder for
energy and momentum to be conserved.

Let us now consider processes in which magnetorotons with wave vectors close
to the minimum gap are scattered to longer wave lengths by absorbing phonons.
In this case we have
\begin{equation}
\Delta\left(  k\right)  \sim\Delta^{\ast}+\frac{\left(  k-q^{\ast}\right)
^{2}}{2\mu}%
\end{equation}
where $\mu$ is an effective magnetoroton mass. In this case the absorption of
a small phonon with in-plane wave vector $q$ will be allowed provided\qquad\
\begin{equation}
\left|  k-q^{\ast}\right|  >\frac{\mu c_{s}}{\sin\theta}%
\end{equation}
this gives, for $\nu=1/3$ (assuming the ideal magnetoroton dispersion)
$\left|  k-q^{\ast}\right|  >0.19/l_{c}$ for LA\ phonons and $\left|
k-q^{\ast}\right|  >0.12/l_{c}$ for TA\ phonons.

\section{Magnetoroton Scattering Contribution to Phonon Absorption}

The leading correction to the energy transfer rate due to magnetoroton
creation at finite electron temperature will be due to the processes discussed
above in which a magnetoroton is scattered from one mode to another with the
absorption of a phonon. This contribution to the energy transfer rate will be
given by
\begin{equation}
\delta P=\sum_{s,\underline{Q},\mathbf{k}^{\prime}}\omega_{s}\left(
\underline{Q}\right)  \tau_{\mathbf{k}^{\prime}\mathbf{,s,}\underline{Q}}%
^{-1}n_{B}\left(  \omega_{s}\left(  \underline{Q}\right)  /T_{\phi}\right)
n_{B}\left(  \Delta\left(  k^{\prime}\right)  /T_{e}\right)
\end{equation}
where $T_{\phi}$ is the characteristic temperature of the phonon pulse and
$n_{B}\left(  x\right)  =\left(  e^{x}-1\right)  ^{-1}$ is the Bose
distribution function and we assume that $T_{e}\ll T_{\phi}\ll\Delta^{\ast}$
so that $n_{B}\left(  \Delta\left(  k\right)  /T_{e}\right)  \ll1$ $\forall
k$. In order to simplify matters we will neglect the finite thickness of the
2d layer (i.e. we set $Z\left(  q_{z}\right)  =1$) and neglect the anisotropy
in the electron phonon interaction (which at long wavelengths will be
dominated by the piezoelectric coupling) and set
\begin{equation}
M_{s}\left(  \underline{Q}\right)  =\frac{\Lambda_{s}}{\sqrt{Q}}\qquad
\end{equation}
(strictly this bare piezoelectric coupling will be screened by the equilibrium
population of excitations but we will not attempt to estimate such effects
here). Finally we will use the isotropic Debye approximation (and hence
neglect phonon-focussing effects) setting
\begin{equation}
\omega_{s}\left(  \underline{Q}\right)  =c_{s}Q\qquad.
\end{equation}
In the regime $T_{e}\ll T_{\phi}\ll\Delta^{\ast}$ the transfer of energy will
be dominated by absorption by magnetorotons initially at the minimum,
$k=q^{\ast}\approx1.34/l_{c}$, hence we will have
\begin{align}
\delta P  & \sim\sqrt{2\pi\mu T_{e}}e^{-\Delta^{\ast}/T_{e}}\sum_{s}%
c_{s}\Lambda_{s}^{2}\int\frac{d^{3}\underline{Q}}{8\pi^{3}}\frac{\left|
P^{\left(  3\right)  }\left(  q^{\ast}\widehat{\mathbf{n}}\mathbf{,q}\right)
\right|  ^{2}}{\overline{s}\left(  q^{\ast}\right)  \overline{s}\left(
\left|  q^{\ast}\widehat{\mathbf{n}}\mathbf{+q}\right|  \right)  }%
\delta\left(  \Delta\left(  \left|  q^{\ast}\widehat{\mathbf{n}}%
\mathbf{+q}\right|  \right)  -\Delta^{\ast}-c_{s}Q\right)  \\
& \times n_{B}\left(  c_{s}Q/T_{\phi}\right)  \qquad
\end{align}
where $\widehat{\mathbf{n}}$ is an arbitrary vector in the plane. Following
\cite{Theory} we transform to new variables to integrate out the
delta-function and arrive at
\begin{align}
\delta P  & \sim\sqrt{\frac{\mu T_{e}}{2\pi}}e^{-\Delta^{\ast}/T_{e}}\sum
_{s}\Lambda_{s}^{2}\int\frac{d^{2}\mathbf{k}}{4\pi^{2}}\frac{\omega
_{k}\vartheta\left(  \omega_{k}-c_{s}\left|  \mathbf{k-}q^{\ast}%
\widehat{\mathbf{n}}\right|  \right)  }{\sqrt{\omega_{k}^{2}-c_{s}^{2}\left|
\mathbf{k-}q^{\ast}\widehat{\mathbf{n}}\right|  ^{2}}}\\
& \times\frac{\left|  P^{\left(  3\right)  }\left(  q^{\ast}\widehat
{\mathbf{n}}\mathbf{,k-}q^{\ast}\widehat{\mathbf{n}}\right)  \right|  ^{2}%
}{\overline{s}\left(  q^{\ast}\right)  \overline{s}\left(  k\right)  }%
n_{B}\left(  \omega_{k}/T_{\phi}\right)
\end{align}
where the integration is now over all allowed transitions from the initial
state $\left|  q^{\ast}\widehat{\mathbf{n}}\right\rangle $ to a final state
$\left|  \mathbf{k}\right\rangle $ and
\begin{equation}
\omega_{k}=\Delta\left(  k\right)  -\Delta^{\ast}%
\end{equation}
is the corresponding energy change. The effective transition rate is simply
understood by using a polar representation for the $\mathbf{k}$-plane, with
$\widehat{\mathbf{n}}$ as the polar axis. The scatering rate is then
\begin{align}
\tau_{\mathbf{k}}^{-1}  & =\frac{\vartheta\left(  \omega_{k}^{2}-c_{s}%
^{2}\left(  k^{2}+q^{\ast2}-2q^{\ast}k\cos\phi\right)  \right)  }{\sqrt
{\omega_{k}^{2}-c_{s}^{2}\left(  k^{2}+q^{\ast2}-2q^{\ast}k\cos\phi\right)  }%
}\Phi\left(  k,\phi\right)  \\
\Phi\left(  k,\phi\right)    & =\frac{\left|  P^{\left(  3\right)  }\left(
q^{\ast}\widehat{\mathbf{n}}\mathbf{,k-}q^{\ast}\widehat{\mathbf{n}}\right)
\right|  ^{2}}{\overline{s}\left(  q^{\ast}\right)  \overline{s}\left(
k\right)  }\qquad.
\end{align}
Numerical study, using the form given above for $\Delta\left(  k\right)  $
(equation \ref{Dispersion}) and the speed of sound, $c_{s}$, appropriate for
TA\ phonons, shows that the condition imposed by the theta-function is only
satisfied for final states within a contour $k_{c}\left(  \phi\right)  $
(shown in figure 2) in the $\mathbf{k}$-plane, confirming the simple analysis
presented in the previous section. Only the states inside this contour are
accessible from the initial state $\left|  q^{\ast}\widehat{\mathbf{n}%
}\right\rangle $. Along any direction in the $\mathbf{k}$-plane, there is a
square root singularity as $k\rightarrow k_{c}\left(  \phi\right)  $ which
will dominate the rate of energy transfer. The form of $\tau^{-1}$ along
several representative directions in the $\mathbf{k}$-plane are shown in
figure 3. In this case we have used the simple ansatz for $P^{\left(
3\right)  }$ proposed by MacDonald et al.\cite{MacDonald}
\begin{equation}
P^{\left(  3\right)  }\left(  \mathbf{k,q}\right)  \approx2i\sin\left(
\frac{1}{2}\left|  \mathbf{k\times q}\right|  \right)
\end{equation}
although results obtained from the convolution approximation (equations
\ref{P3}\ and \ref{Convolution}) are qualitatively similar.\ In the appendix
below, the behaviour of $\delta P$ is considered further and it is shown that
for a given $\phi$ there is a characteristic temperature $T_{X}\left(
\phi\right)  =\Delta\left(  k\left(  \phi\right)  \right)  -\Delta^{\ast}$
above which the dependence on heater temperature, $T_{\phi}$, develops a
non-activated component.

It is clear then, that a non-activated component will appear in $\delta P$
when $T_{\phi}>\inf_{\phi}\left(  T_{X}\left(  \phi\right)  \right)
=T_{X}\left(  0\right)  $. To give an idea of the physical orders of magnitude
involved, consider a 2des with sheet density $n_{s}=10^{15}%
\operatorname{m}%
^{-2}$. The $\nu=1/3$ fractional quantum Hall state will occur at a field
$B=12.4$T. The cyclotron length is then $l_{c}=\sqrt{\hbar/eB}=7.3%
\operatorname{nm}%
$ and the Coulomb energy scale is $\mathcal{E}_{c}=e^{2}/4\pi\epsilon
_{0}\kappa l_{c}=15.2$m$%
\operatorname{eV}%
$ (assuming $\kappa=13$ appropriate for GaAs). Using the form of the
magnetoroton dispersion derived by GMP \cite{GMP} gives $\Delta^{\ast}%
\simeq0.075\mathcal{E}_{c}=1.14$m$%
\operatorname{eV}%
$ which corresponds to a temperature of $13$K and $q^{\ast}\simeq1.3l_{c}%
^{-1}=$ $1.8\times10^{8}%
\operatorname{m}%
^{-1}$. Using this dispersion relation we find, for TA\ phonons, $k\left(
0\right)  =1.08/l_{c}$ which gives $T_{X}\left(  0\right)  =\Delta\left(
k\left(  0\right)  \right)  -\Delta^{\ast}=0.005\mathcal{E}_{c}=0.067\Delta
^{\ast}$ corresponding to a temperature of $870$mK. Given that the value of
the gap extracted from the phonon experiments (which is reduced from the ideal
value by the combined effects of disorder, finite thickness of the 2des and
Landau level mixing) is more like $4$K \ we might expect $T_{X}\left(
0\right)  $%
$<$%
$200$mK: well below the heater temperature used in any of the experiments (and
close to the base temperature of the dilution refrigerator) . Hence we deduce
that there will always be a contribution to the energy transfer rate which is
non-exponential in the heater temperature (although its dependence of the
electron temperature is still of the form $e^{-\Delta^{\ast}/T_{e}}$.

\section{Conclusions}

In this paper we have considered processes in which magnetorotons existing at
finite electron temperature $T_{e}$ are scattered by absorbing phonons from a
pulse characterized by a phonon temperature $T_{\phi}$. We have shown that,
contrary to our earlier supposition \cite{Wurtzburg}, such processes will not
heat the magnetorotons to sufficiently large wave vectors that they dissociate
into separate charged quasiparticles and so directly contribute to the
longitudinal conductivity. This negative result is important in that the
phonon absorption experiments \cite{PRL1}\cite{PRL2} measure the change in
longitudinal conductivity as a ballistic phonon pulse hits the 2des. The
absence of any direct dissociation mechanisms means that the energy
transferred to the 2des is thermalized within the 2des, raising its effective
temperature and hence leading to an increase in the thermal population of
charged quasiparticles. Consequently, the changes in $\sigma_{xx}$ can be
calibrated against equilibrium magneto-transport experiments to give a direct
measure of the non-equilibrium electron temperature.

Although the processes discussed can not dissociate the magnetorotons into
charged quasiparticles, scattering to longer wavelengths is possible and will
give a contribution to the energy transfer rate which is exponential in the
electron temperature but will have an non-exponential form except at the very
lowest heater temperatures. Such a power law dependence (on $T-T_{X}\left(
0\right)  $) might well be observable in current phonon absorption experiments
performed over a range of base electron temperatures. It has recently been
shown \cite{Jain} that the lowest lying neutral excitation at long wavelengths
is not the small $q$ magnetoroton but a bound state of two magnetorotons. It
is then possible that the absorption of energy from a higher temperature
phonon pulse may involve a cascade process in which magnetorotons are
scattered to long wavelengths by absorption of low energy phonons and then
decay into two roton bound states which are thermally dissociated, leading to
an enhanced population of magnetorotons at the minimum.

\begin{acknowledgement}
This work was supported by the EPSRC (UK). We would like to acknowledge many
useful discussions on this topic with CJ\ Mellor, U Zeitler, S. Roshko, A.
Devitt and J. Digby.
\end{acknowledgement}

\appendix

\section{Analysis of energy transfer rate}

We have the following form for the scattering contribution to the energy
transfer rate
\begin{equation}
\delta P\sim\frac{1}{4\pi^{2}}\sqrt{\frac{\mu T_{e}}{2\pi}}e^{-\Delta^{\ast
}/T_{e}}\sum_{s}\Lambda_{s}^{2}\int_{0}^{2\pi}d\phi Y\left(  \phi\right)
\int_{0}^{k\left(  \phi\right)  }dk\frac{\vartheta\left(  \omega_{\mathbf{k}%
}-c_{s}\left|  \mathbf{k-}q^{\ast}\widehat{\mathbf{n}}\right|  \right)
}{\sqrt{\omega_{k}^{2}-c_{s}^{2}\left|  \mathbf{k-}q^{\ast}\widehat
{\mathbf{n}}\right|  ^{2}}}n_{B}\left(  \omega_{k}/T_{\phi}\right)
\end{equation}
where
\begin{equation}
Y\left(  \phi\right)  =k\left(  \phi\right)  \omega\left(  k\left(
\phi\right)  \right)  \frac{\left|  P^{\left(  3\right)  }\left(  q^{\ast
}\widehat{\mathbf{n}}\mathbf{,k}\left(  \phi\right)  \mathbf{-}q^{\ast
}\widehat{\mathbf{n}}\right)  \right|  ^{2}}{\overline{s}\left(  q^{\ast
}\right)  \overline{s}\left(  k\left(  \phi\right)  \right)  }%
\end{equation}
and
\begin{equation}
\mathbf{k}\left(  \phi\right)  =k\left(  \phi\right)  \left(  \cos\phi
\widehat{\mathbf{n}}+\sin\phi\widehat{\mathbf{\phi}}\right)
\end{equation}
so that
\begin{align}
\delta P  & \sim\frac{1}{4\pi^{2}}\sqrt{\frac{\mu T_{e}}{2\pi}}e^{-\Delta
^{\ast}/T_{e}}\sum_{s}\Lambda_{s}^{2}\int_{0}^{2\pi}d\phi\frac{Y\left(
\phi\right)  }{\sqrt{\left|  Q\left(  \phi\right)  \right|  }}\\
& \times\int_{0}^{k\left(  \phi\right)  }\frac{dk}{\sqrt{k\left(  \phi\right)
-k}}n_{B}\left(  \frac{\Delta\left(  k\left(  \phi\right)  \right)
-\Delta^{\ast}+\left(  k\left(  \phi\right)  -k\right)  \left|  \Delta
^{\prime}\left(  k\left(  \phi\right)  \right)  \right|  }{T_{\phi}}\right)
\end{align}
where
\begin{equation}
Q\left(  \phi\right)  =2\left[  c_{s}^{2}\left(  k\left(  \phi\right)
-q^{\ast}\cos\phi\right)  -\Delta^{\prime}\left(  k\left(  \phi\right)
\right)  \left(  \Delta\left(  k\left(  \phi\right)  \right)  -\Delta^{\ast
}\right)  \right]  \qquad.
\end{equation}
Now we define
\begin{align*}
a\left(  \phi\right)    & =\Delta\left(  k\left(  \phi\right)  \right)
-\Delta^{\ast}\\
b\left(  \phi\right)    & =\left|  \Delta^{\prime}\left(  k\left(
\phi\right)  \right)  \right|  \\
\tau & =T_{\phi}/a\left(  \phi\right)
\end{align*}
so that we need to consider the integral
\begin{equation}
\int_{0}^{\infty}\frac{d\eta}{\sqrt{\eta}}\frac{1}{e^{\left(  a+b\eta\right)
/a\tau}-1}=\frac{T_{\phi}}{\sqrt{ab}}\int_{1/\tau}^{\infty}\frac{dx}%
{\sqrt{\tau x-1}\left(  e^{x}-1\right)  }\qquad.
\end{equation}
We will use the an interpolation formula for the Bose function
\[
n_{B}\left(  x\right)  =\left\{
\begin{array}
[c]{cc}%
1/x & x<1\\
e^{-x} & x>1
\end{array}
\right.
\]
and consider two cases. Firstly, in the case that $\tau<1$ we have that
\begin{align}
\int_{1/\tau}^{\infty}\frac{dx}{\sqrt{\tau x-1}\left(  e^{x}-1\right)  }  &
\sim\int_{1/\tau}^{\infty}\frac{e^{-x}dx}{\sqrt{\tau x-1}}\nonumber\\
& =\sqrt{\frac{\pi}{\tau}}e^{-1/\tau}\qquad.
\end{align}
In the other case, we have
\begin{align}
\int_{1/\tau}^{\infty}\frac{dx}{\sqrt{\tau x-1}\left(  e^{x}-1\right)  }  &
\sim\int_{1/\tau}^{1}\frac{dx}{x\sqrt{\tau x-1}}+\int_{1}^{\infty}\frac
{e^{-x}dx}{\sqrt{\tau x-1}}\nonumber\\
& =2\arctan\left(  \sqrt{\tau-1}\right)  +\sqrt{\frac{\pi}{\tau}}e^{-1/\tau
}\left(  1-\operatorname{erf}\left(  \frac{\sqrt{\tau-1}}{\sqrt{\tau}}\right)
\right)
\end{align}
Hence, for $\tau>1$ there is a contribution that is non-activated in the
phonon temperature. The angular ($\phi$) integration is non-trivial and its
numerical evaluation is rather pointless given our neglect of anisotropy in
phonon propagation and the electron-phonon coupling. It would however be
expected, for $T_{\phi}\gtrsim T_{X}\left(  0\right)  $ to have a power law
dependence on $T_{\phi}-T_{X}\left(  0\right)  $.

{\Large Figure Captions}

Figure 1: A plot of form of the magnetoroton dispersion $\Delta\left(
\mathbf{q}\right)  $ used here.

Figure 2: A plot of the contour $k_{c}\left(  \phi\right)  $ in $\mathbf{k}%
$-space. Only magnetoroton states in the shaded region inside the contour are
accessible from the initial state $\left|  q^{\ast}\widehat{\mathbf{n}%
}\right\rangle $ via a single phonon scattering process.

Figure 3: Representative plots of the scattering rate into states with
wavenumber $k$ for a sample set of directions, $\phi$. The square root
singularities in each trace occur at the critical value $k_{c}\left(
\phi\right)  $.

\begin{thebibliography}{99}
\bibitem{gen-phon}L.J. Challis and A.J. Kent, in ``Aspects of Semiconductor
Nanostructures'' (ed. G. Bauer, F. Kuchar and H. Heinrich) (Heidelberg:
Springer, 1992) p31.

\bibitem {PRL1}C.J.\ Mellor, R.H. Eyles, J.E. Digby, A.J. Kent, K.A. Benedict,
L.J. Challis, M.\ Henini, C.T.\ Foxon, Phys. Rev.\ Lett. \textbf{74}, 2339 (1995).

\bibitem {PRL2}U. Zeitler, A.M. Devitt, J.E. Digby, C.J. Mellor, A.J. Kent,
K.A. Benedict, T. Cheng, Phys. Rev. Lett., \textbf{82}, 5333 (1999).

\bibitem {GMP}S.M. Girvin, A.H. MacDonald and P.M. Platzman, Phys Rev.
\textbf{B33}, 2481 (1986).

\bibitem {Theory}K.A. Benedict, R.K. Hills, C.J. Mellor, Phys. Rev. B
\textbf{60}, 10984 (1999).

\bibitem {Platzman}P.M. Platzman, Phys Rev \textbf{B39}, 7985 (1989).

\bibitem {Wurtzburg}K.A. Benedict in ``Proceedings of the Twelfth
International Conference on High Magnetic Fields in the Physics of
Semiconductors'', (Singapore: World Scientific), 95 (1997).

\bibitem {Laughlin}R.B. Laughlin, Phys Rev Lett. \textbf{50}, 1395 (1983).

\bibitem {Feynman}R.P.\ Feynman, Phys Rev \textbf{91}, 1291 (1953).

\bibitem {Jachs}S.M. Girvin and T. Jach, Phys Rev B \textbf{29}, 5617 (1984).

\bibitem {MacDonald}A.H. MacDonald, K.L. Liu, S.M. Girvin, P.M. Platzman,
Phys. Rev. \textbf{B33}, 4014 (1986).

\bibitem {Jackson}H.W.\ Jackson and E. Feenberg, Rev. Mod. Phys. \textbf{34},
686 (1962).

\bibitem {Brownlie}M. Brownlie and K.A. Benedict, J. Phys. A. \textbf{33},
4283\emph{\ }(2000).

\bibitem {Jain}K. Park and J.K. Jain, Phys. Rev. Lett. \textbf{84}, 5576 (2000).
\end{thebibliography}
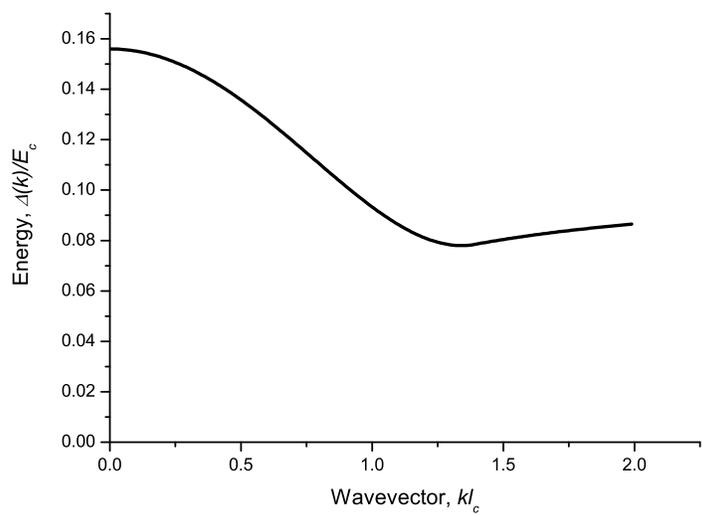
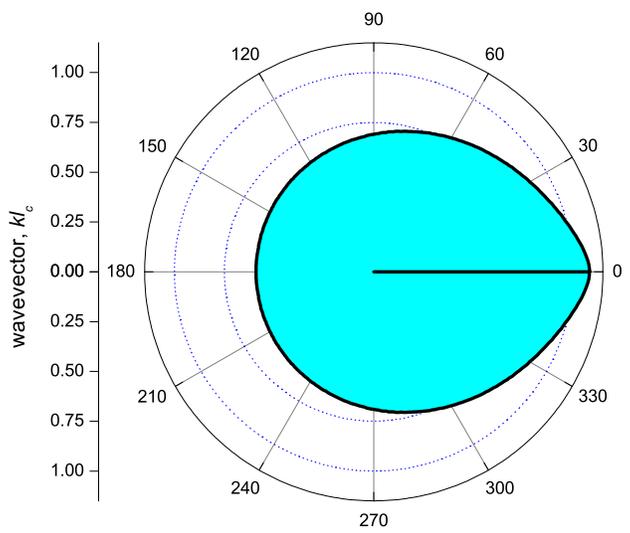
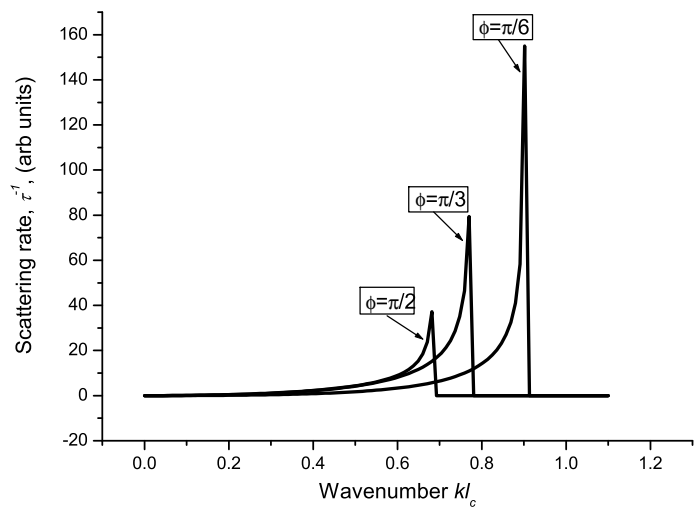
\end{document}